\documentstyle[12pt,amssymbols]{article}
\newtheorem{Lemma}{Lemma}[section]
\newtheorem{Theorem}{Theorem}[section]
\newtheorem{Corollary}{Corollary}[section]

\newcommand{\beq}{\begin{equation}}
\newcommand{\eeq}{\end{equation}}
\newcommand{\bq}{\begin{quotation}}
\newcommand{\eq}{\end{quotation}}
\newcommand{\bc}{\begin{center}}
\newcommand{\ec}{\end{center}}

\newcommand{\tr}   {\mbox{tr}}
\newcommand{\Cl}   {\mbox{Cl}}
\newcommand{\Int}   {\mbox{Int}}
\newcommand{\Real}  {\mbox{Re}}

\newcommand{\Mzer} {{{\cal M}_0}}
\newcommand{\Mone} {{{\cal M}_1}}

\newcommand{\half} {\frac{1}{2}}
\newcommand{\rr}    {\rightarrow}

\newcommand{\bo}{{\bar{\omega}}}
\newcommand{\mysubsect}[1]  {\subsection{\it{#1} } }

\newcommand{\CC}{\mbox{$\Bbb C$}}
\newcommand{\RR}{\mbox{$\Bbb R$}}

\newcommand{\diagram}[9]{
  \beq
  \begin{array}{ccccccc}
   \multicolumn{3}{c}{#1} & \stackrel{#2}{\longr} & \multicolumn{3}{c}{#3}\\
   #4   & \downarrow      &       &  &            & \downarrow  &      #5 \\
   \multicolumn{3}{c}{#6} & \stackrel{#7}{\longr} & \multicolumn{3}{c}{#8}\\
  \end{array}
  \label{#9}
  \eeq
  }

\newcommand{\Sp}{\mbox{sp}}
\newcommand{\dist}{\mbox{dist}}
\newcommand{\lap}{\int_0^\infty \!\!dt\;}
\newcommand{\Halmos}    {\raisebox{0.5ex}  {\framebox[1.1ex]{
                              \rule[0ex]{0ex}{1ex}
                               }}}
\newcommand{\longr}{\longrightarrow}

\renewcommand{\baselinestretch}{1.3}

\title{Intermittency and Regularized Fredholm Determinants}
\author{Hans Henrik Rugh \\
       (University of Warwick, Coventry CV4 7AL, UK)}
\date {\today}
\begin{document}
\maketitle
 \begin{abstract}
 We consider real-analytic maps of the interval $I=[0,1]$ which are expanding
 everywhere except for a neutral fixed point at $0$. We show that on a certain
 function space the spectrum of the associated Perron-Frobenius operator
 ${\cal M}$ has a decomposition $\Sp({\cal M}) = \sigma_c \cup \sigma_p$
 where $\sigma_c=[0,1]$ is the continuous spectrum of ${\cal M}$ and
 $\sigma_p$ is the pure point spectrum with no points of accumulation outside
 $0$ and $1$. We construct a regularized Fredholm determinant $d(\lambda)$
 which has a holomorphic extension to $\lambda \in \CC-\sigma_c$ and
 can be analytically continued from each side of $\sigma_c$ to an open
 neighborhood of $\sigma_c-\{0,1\}$ (on different Riemann sheets). In
 $\CC-\sigma_c$ the zero-set of $d(\lambda)$  is in one-to-one correspondence
 with the point spectrum of ${\cal M}$.  Through the conformal transformation
 $\lambda(z) = \frac{1}{4z} (1+z)^2$ the function $d \circ \lambda(z)$ extends
 to a holomorphic function in a domain which contains the unit disc.\\

Shorttitle : Intermittency and Regularized Fredholm Determinants.
 \end{abstract}

\section{Assumptions and statement of results.}
Finding analytic continuations of functions, holomorphic
in some a priori given domains, is a challenging and rewarding
mathematical task in itself.  In some cases it also
provides elegant and non-trivial solutions
to other problems  in mathematics or  physics.
Thus, in dynamical systems theory  the
spectral properties of transfer operators
and the zero sets
of analytically extended holomorphic  functions 
are related through Ruelle's
generalized Fredholm determinants and dynamical zeta functions.
In establishing such a relationship uniform
hyperbolicity traditionally plays a crucial role and 
have led to interesting  results e.g. in the cases of 
purely expanding maps or hyperbolic flows,
both on compact manifolds and  with
various degrees of smoothness imposed 
\cite{Ruelle76,Ruelle90}. In these cases, 
an associated Perron-Frobenius type 
 positive bounded transfer operator ${\cal M}$ has a spectral
radius $r({\cal M})$, and an essential spectral radius\footnote
        {The spectral `rug' under which you sweep the part of
         the spectrum you don't understand.}
$r_{\rm ess}({\cal M})$ strictly smaller than   $r({\cal M})$.
The part of the spectrum which intersects
the annulus 
$(r_{\rm ess},r({\cal M})]$ is then non-empty and
isomorphic to the zero-set
of a Fredholm-determinant.
 This in turn yields
informations about the decay of correlations and
in some cases geometrical properties of the underlying manifold
\cite{PP}.
However, in the presence of neutral fixed points,
e.g. \cite{PM} of the Pomeau-Manneville intermittency type 1, the
two spectral radii coincide and the annulus, whence
the result about the spectrum, becomes void.
Recently, using the technique of induced mappings, 
Prellberg \cite{PS92} and also Isola \cite{Isola95}
have overcome part of this problem and 
obtained some interesting spectral properties.
We refer also to Mayer \cite{Mayer91} for a closely related
analysis of the Gauss map.\\

Below we shall develop a theory which combines
results from complex dynamics, for which we rely on the exposition
of Shishikura \cite{Shishikura}, with the nuclear theory
of Grothendieck \cite{Grot55,Grot56} and  the 
dynamical zeta-function analysis first
introduced by Ruelle \cite{Ruelle76}.
We shall consider the intermittent case for
certain real-analytic  mappings
of an interval and  associate to such a map a Perron-Frobenius type
transfer operator ${\cal M}$.  We introduce
a regularized Fredholm determinant, $d(\lambda)$, and consider the relationship
between its analytic properties and the spectrum of ${\cal M}$.
By first using spectral properties of ${\cal M}$ we deduce
that $d(\lambda)$ is analytic in the complement
of the line segment $\sigma_c=[0,1]$ (where eigenvalues of ${\cal M}$
can be identified with zeroes of $d(\lambda)$)
 and then by analytically
continuing  $d$ 
 across the open segment $(0,1)$ from both sides
(onto different Riemann sheets)
 we deduce that
the pure point spectrum of
${\cal M}$ can not have points of accumulation apart from
the end points 0 and 1.
 Moreover, we find  through a conformal 
transformation an expression for the Fredholm determinant,
accessible to numerical analyses. These results are special cases
(section \ref{dyn-sys})
of a slightly more general set-up which we shall now describe.\\

We consider an analytic map $f : \Delta \rr \Delta$ of an
open, simply connected domain $\Delta \subset \CC^*=\CC-\{0\}$.
Also, let $g$ be an analytic function on $\Delta$.
We shall assume that both $f$ and $g$ have continuous extensions to
the boundary of $\Delta$. For $r_0>0$, $\pi > \theta_0>0$, the set
${\cal S}[r_0,\theta_0] = \{re^{i\theta} \in \CC :
     r \in (0,r_0), \theta \in (-\theta_0,\theta_0) \}$ will denote an
open angular sector.\\

We say that  $f$ is 
an intermittent analytic contraction at $0$ of order $1+p > 1$
and that $g$ is an associated weight provided~:\\

\indent (i)  $f : \Cl \; \Delta \rr  \Delta \cup \{0\}$ and $f$ is univalent 
            in $\Delta$.\\[2mm]
\indent (ii)  There is $r_0>0$, $\theta_0 > \frac{\pi}{2p}$
 such that $\Delta$ contains
    the open angular sector ${\cal S}[r_0,\theta_0]$.\\[2mm]
\indent (iii)  There are constants $\epsilon>0, a>0$ and $b\in \CC$ such that
       for $z \in \Delta$

         \mbox{} \ \  $f(z)  = z - a z^{1+p} + {\cal O}(|z|^{1+p+\epsilon})$ ,

         \mbox{} \ \  $g(z)  = 1 - b z^p + {\cal O}(|z|^{p+\epsilon})$ .\\

In the following let
$\mu(d\omega)$ be a finite, positive
measure on a measure space
$\Xi$. With respect to this measure
let $\{f_\omega : \Delta \rr \Delta\}_{\omega \in \Xi}$
 be a measurable family of analytic maps with continuous
extension to the boundary. We shall assume that the family
is uniformly contracting in $\Delta$ by which we mean that\\[2mm]
\indent (iv) $ \Cl \ \bigcup_\omega f_\omega (\Delta) \subset \Delta$.\\[2mm]
\noindent Also, let $\{g_\omega : \Delta \rr \CC\}_{\omega \in \Xi}$
be a measurable family of analytic (weight-) functions
with continuous extension to the boundary and such that\\[2mm]
\indent (v)
$ \int_\Xi \mu(d\omega) \sup_{z \in \Delta}
 |g_\omega(z)| < \infty $.\\[2mm]
For notational convenience
we introduce $\Xi^* = \Xi \cup \{0\}$ and extend $\mu$ to $\Xi^*$
by setting $\mu^*(\{0\}) = 1$. We set $f_0=f$ and $g_0=g$
and define $\Sigma^* \subset  \Delta \cup \{0\}$ to be the compact 
invariant set under the extended family $\{f_\omega\}_{\omega\in\Xi^*}$.
 We define for each $\omega \in \Xi^*$  an operator
  ${\cal M}_\omega   :   {\cal H} (\Delta) \rr  {\cal H}(\Delta)$
acting 
on functions analytic in $\Delta$ (cf. section \ref{section-proof}
for the definition of ${\cal H}(\cdot)$),
\begin{equation}
  \label{eq-PFfirst} 
  {\cal M}_\omega h(z) = 
             g_\omega(z) h \circ f_\omega(z) 
\end{equation}
and 
 we set
  $ {\cal M} = {\cal M}_0 + {\cal M}_1 $ where
\begin{equation}
  {\cal M}_1 = \int_{\Xi}
            \mu(d\omega) {\cal M}_\omega , 
  \label{eq-PFlast}
\end{equation}
which is well-defined by properties (iv)-(v).
\noindent
For $\bar{\omega} = (\omega_1,...,\omega_m)\in (\Xi^*)^m$, $m>0$
we introduce the abbreviations~:
\begin{eqnarray}
 f_{\bar{\omega}}(z) &=&
     f_{\omega_1} \circ \cdots \circ f_{\omega_m} (z) \\
 g_{\bar{\omega}}(z) &=&
     g_{\omega_m}(z) 
     g_{\omega_{m-1}}(f_{\omega_m}z)\cdots
     g_{\omega_1} (f_{\omega_2,...,\omega_m}z)
\end{eqnarray}
and we let $x(\bar{\omega})$ 
denote the (necessarily unique) fixed point of $f_{\bar{\omega}}$
in $\Cl \; \Delta$. The regularized m'th direct product of the
family $\Xi^*$ consists of
$\Xi^*_m = (\Xi^*)^m - \{0 \times \cdots \times 0\}$, i.e.
the direct product minus
the element consisting of $m$ repeats of the
intermittent fixed point. We write
\begin{eqnarray}
  \zeta_m&=&  \int_{(\Xi^*)^m} \mu^*(d \omega_1)\cdots \mu^*(d \omega_m)
          { g_{\bar{\omega}}(x(\bar{\omega}))} ,
          \label{eq:zetacoef}\\
  d_m &=&  \int_{\Xi^*_m} \mu^*(d \omega_1)\cdots \mu^*(d \omega_m)
          \frac{ g_{\bar{\omega}}(x(\bar{\omega}))}
          {1-f'_{\bar{\omega}}(x(\bar{\omega}))} .
          \label{eq:fredcoef}
\end{eqnarray}
The first integral thus extends over all possible periodic orbit
of length $m$
of the extended family whereas in the latter the
intermittent fixed point itself (and only that) has been excluded.
A zeta function and a regularized Fredholm determinant is then defined through
the following formal power series in $z$ and $\lambda^{-1}$
respectively~:

\begin{eqnarray}
 \zeta (z) &=& \exp  \sum_{m=1}^\infty \frac{z^{m}}{m} \zeta_m .
 \label{eq:zeta}\\
 d(\lambda) &=& \exp - \sum_{m=1}^\infty \frac{\lambda^{-m}}{m} d_m .
 \label{eq:fredholm}
\end{eqnarray}
\begin{Theorem}
Assume that our collection of maps and weights satisfy
properties (i) through (v).
If also  $g \neq 0$ on $\Sigma^*$ and Re$(b) > pa$ (the constants as 
in  property (iii))
then there is a domain $U \subset \Delta$ which contains the invariant
set $\Sigma^*-\{0\}$ and a
Banach space $X(U) \subset {\cal H}(U)$ of analytic functions in $U$
such that (\ref{eq-PFfirst}-\ref{eq-PFlast}) defines a bounded linear operator
 ${\cal M} : X(U) \rr X(U)$. Furthermore,
\begin{itemize}
\item[(a)]
 The spectrum of ${\cal M}$ has a decomposition
 $\Sp({\cal M}) = \sigma_c \cup \sigma_p$ where $\sigma_c=[0,1]$
   (the line-segment)
   is the continuous spectrum of ${\cal M}$
   and $\sigma_p$ is
    the pure point
   spectrum of ${\cal M}$.
\item[(b)] $\sigma_p$ consists of eigenvalues of finite multiplicity
    and has no points of accumulations in
    $\CC - (\{0\} \cup \{1\})$.
\item[(c)] 
   $d(\lambda)$ extends to a holomorphic function in $\bar{\CC}-\sigma_c$
 where its zeroes counted with order are the same as the eigenvalues
of ${\cal M}$ counted with multiplicity. 
\item[(d)]
   $d(\lambda)$ has an analytic continuation from each side of 
   $\sigma_c$ to 
   an open neighborhood of 
   $\sigma_c - \{0,1\}$.
\end{itemize}
\label{first-theorem}
\label{WHY-CAPITAL-LETTERS}
\end{Theorem}

In the following Corollaries 
\ref{conformal-corollary} through
\ref{zeta-corollary}
 we shall assume that the conditions of the Theorem
are satisfied.

\begin{Corollary}
Through the conformal transformation
  $ \lambda(z) = \frac{1}{4z}(1+z)^2$
the function
\begin{equation}
  d(\lambda(z)) = \exp - \sum_{m>0} \frac{z^m}{m}
       \left( \frac{2}{1+z} \right)^{2m} d_m
\end{equation}
extends to a holomorphic function in a domain which contains
 the open unit disc. Here
its zero-set is in one-to-one correspondence with the 
point spectrum of ${\cal M}$
in $\CC-\sigma_c$.
\label{conformal-corollary}
\end{Corollary}

Standard results from nuclear theory yields that
the Fredholm determinant $d$ is holomorphic in $\CC$ minus the unit disc
(cf. \cite{PS92}) and hence, that
$d\circ \lambda$  is holomorphic in
 the disc of radius
$3-\sqrt{8} = 0.17...$ The non-trivial content
of the Theorem and its Corollary is that the  analyticity extends to
$\CC-\sigma_c$, respectively to  the open unit-disc.
We remark, that
the function $d\circ \lambda$ is accessible for numerical analysis
through formal power series.\\

\begin{Corollary} (to the proof of Theorem \ref{first-theorem})
\label{extension-corollary}
Let $\nu = p \theta_0  - \frac{\pi}{2}$ $(\in (0,\frac{\pi}{2}))$.
The Fredholm determinant $d(\lambda)$
has an analytic continuation 
to a Riemann surface which is constructed in the following way.
Take the slit Riemann sphere, $\bar{\CC}-\sigma_c$, and cut it
open along the open line segment $(0,1)$. To each side of this segment
glue a half-infinite logarithmic
spiral, turning clockwise, respectively anti-clockwise
 around the origin and with the radius, r, decreasing
as a function of the angle, $\phi$, as
$r(\phi) =\exp ( -\cot(\nu) |\phi|)$.
When $\nu=\pi/2$ (and thus $r(\phi) \equiv 1$)
, this surface is the same as
the one obtained by taking
the universal cover of the open unit disc  punctured at the
origin $D-\{0\}$, pick one of the copies (cut from
0 to 1) and 
 gluing to its circumference (the unit circle minus the point $\{1\}$)
the rest of the Riemann sphere (minus $\{1\}$).
\end{Corollary}

For the dynamical zeta-function associated with our
intermittent system we have~:

\begin{Corollary}
Assume that the derivatives $\{f'_w\}_{w\in \Xi}$ are uniformly
bounded in $\Delta$.
Then the function $\zeta(z)$ extends to a meromorphic function
in $\CC - [1,\infty)$ and has a meromorphic extension
from both sides to an open neighborhood of the segment $(1,\infty)$.
\label{zeta-corollary}
\end{Corollary}

\mysubsect{Dynamical systems.}

\label{dyn-sys}
Let $F : I\rr I$ be a piece-wise expanding
and real-analytic  map of the
interval $I=[0,1]$ 
and such that each of its branches is surjective on $I$.
Let $\Delta\supset I$ be a complex domain to be specified
in the following.
For simplicity we shall assume that there is only a
finite number $(N+1)$ of branches
and that each inverse  branch, $f_\alpha$,
 $\alpha \in \Xi^*= \{0\} \cup \Xi = \{0,1,...,N\}$
 extends to
an analytic function
 with no critical points in the domain $\Delta$.
Let $\beta \in \CC$ be a
complex parameter and 
let $s_\alpha \in \{1,-1\}$ be the sign of $f'_\alpha$ restricted
to the real interval $I$.
We choose the particular (well-defined) family of
weights $g_\alpha =  (s_\alpha f'_\alpha)^\beta$, $\alpha \in\Xi^*$.
 The domain $\Delta$ should be chosen so that
$f=f_0$ and the weight $g=g_0$  has the
properties (i)-(iii) above    for some values of
the constants $p,r_0,\theta_0,a,b$ and $\epsilon$ and
also that the other branches and weights
$\{f_\alpha,g_\alpha\}_{\alpha \neq 0}$
satisfies (iv) and (v)
($\mu$ is then just the counting measure).\\

 One verifies that the hypotheses of
the Theorem are satisfied provided
Re$\;\beta > \frac{p}{1+p}$.\\
Remark : the above can easily be extended to a numerable
 family of inverse branches, possibly by sharpening
 the condition on $\beta$.\\

 The associated Perron-Frobenius operator
\begin{equation}
{\cal P}_\beta h(z) = \sum_{\alpha}
 (s_\alpha f'_\alpha(z))^\beta h \circ f_\alpha(z)
\end{equation}
defines a continuous operator on ${\cal H}(\Delta)$.
Setting
\begin{eqnarray}
 \zeta_m &=& \sum_{x \in {\rm Fix} F^m}
        {|DF^m(x)|^{-\beta}} ,\\
 d_m &=& \sum_{x \in {\rm Fix} F^m - \{0\}}
        \frac{|DF^m(x)|^{1-\beta}}{|DF^m(x)-1|} ,
\end{eqnarray}
the Theorem with its Corollaries (including \ref{extension-corollary})
apply to the operator ${\cal P}_\beta$, the
associated
 regularized  Fredholm determinant (\ref{eq:fredholm})
 and the zeta-function (\ref{eq:zeta}).

\noindent Example :
The Farey map is defined by~:
\begin{equation}
   F(x) = \left\{
            \begin{array}{cc}
                {x}/{(1-x)} & \mbox{for $x\in (0,\half)$} \\
                {(1-x)}/{x} & \mbox{for $x\in (\half,1)$}
           \end{array} .
          \right.
\end{equation}
For the inverse map $f$ of the left-most branch
we have $1+p=2$ (a parabolic fixed point at 0),
$a=1$ and $\epsilon=1$. The basin of attraction
contains an open angular sector  for any $\theta_0 < \pi$
and we may find a domain $\Delta$ 
satisfying the properties (i)-(iii) and such that
the inverse of the right-most branch is a strict contraction
of $\Delta$ (yielding property (iv)). A calculation shows that
also property (v) is verified.
The Farey map has the Gauss map as induced map.
We refer to Prellberg \cite{PS92} for related results
(on induced maps)
and to Mayer \cite{Mayer91} for a beautiful analysis
of the Gauss map.
\\

\noindent
Acknowledgements : The author would like to thank Viviane Baladi
and the hospitality of the University of Geneva, where this research
was initiated.

\section{Proof of Theorem \protect{\ref{WHY-CAPITAL-LETTERS}}}
\label{section-proof}

For an open  connected domain  $\Omega \subset \CC$ let ${\cal H}(\Omega)$
be the Fr\'echet space of analytic functions in $\Omega$
with the topology generated by the family of sup-norms on
 compact subsets of $\Omega$. For $K \subset \Omega$ closed
and $f \in {\cal H}(\Omega)$ we set $p_K(f) = \sup_{z \in K} |f(z)|$.
By $A(\Omega) = {\cal H}(\Omega) \cap C^0(\bar{\Omega})$
we denote the subset of
analytic functions in $\Omega$ having a continuous
extension to the boundary. $A(\Omega)$ is a Banach space
with the norm given by the supremum norm on $\Omega$. 

For $D$ a subset of 
$\Omega$ we let $r_D$ (omitting the
explicit reference to $\Omega$) denote the restriction map.
The continuity
properties with respect to the underlying function spaces
will be clear from the context.

For any real number $a$ 
we let $H_a = \{ w \in \CC : \Real \; w > a \}$ 
denote the open half plane
of complex numbers
with real part greater than a. \\

We shall first consider the parabolic case where $p+1=2$
and with the special choice of weight $g = (f')^\beta$ with $\beta\in \CC$
 (a restriction on $\beta$ is specified below).
A reduction to this case is given in section \ref{reduction}.

\mysubsect{Complex dynamics and Fatou coordinates. The parabolic case.}
\label{Fatou}

The conditions on $f$ implies that its dynamics may be described
through Fatou coordinates. Let $s (w) = w+1$, $w\in \CC$ be the translation map.
 Using properties (i)-(iii) of the
intermittent contraction
we have the following

\begin{Lemma}
There is an open simply connected domain $\Omega\subset \CC$
 and an analytic bijection $\phi : \Omega \rr \Delta$
such that $\phi$ conjugates
$f : \Delta \rr \Delta$  with the
(well-defined) translation $s : \Omega \rr \Omega$~:
\beq f \circ \phi (w) = \phi (w+1) , \ w \in \Omega .
\eeq
For any $\alpha<\theta_0$, with $\theta_0>\frac{\pi}{2}$ as
in property (ii), there is  $\rho_\alpha \in \RR$ such that
$\Omega$ contains a translated angular sector
$S_\alpha$ of the form
\beq \{\rho_\alpha + r e^{i\theta} : r>0, |\theta|<\alpha \} . 
\label{eq:sector}
\eeq
In any such sector, the functions $\phi'(w)$ and $w^2 \phi'(w)$ are
uniformly bounded.
\label{conjug-lemma}
\end{Lemma}

\noindent proof :

The map $f$ is a self-map of $\Delta$
and it fixes the boundary point $0 \in \partial \Delta$.
It also attracts at least one interior point  (just choose a point
on the positive real axis close to $0$). As a consequence
every
point $z \in \Delta$ converges to $0$ under iteration by $f$.

[To see this, let $z^*$ be an accumulation point
of the sequence $f^k(z)$, i.e.
there is a subsequence
$\psi_n = f^{k_n}$ such that $\psi_n(z)$
converges to $z^* \in \Cl (\Delta)$.
As  $\psi_n$ is a normal family it has itself
a convergent subsequence. The limit function
$\tilde{\psi}$ is analytic in $\Delta$,
 and thus either open or constant. Since
 $\tilde{\psi}(c) = 0$
 for an interior point $c$,
 $\tilde{\psi}$ is not open and
 is therefore the constant map. In particular,
 $z^* = \tilde{\psi}(z) = \tilde{\psi}(c) = 0$. 
 Thus the sequence $f^k(z)$ has $0$ as its only point of accumulation,
 hence it converges to $0$.]

 From the asymptotic form of $f$ one sees that when $f^k z$
 converges to $0$ the argument of $f^k z$ tends to 0 as well,
 i.e. the convergence happens tangentially to the positive real axis.

Let $\sigma : \bar{\CC} \rr \bar{\CC}$
 be the involutory conformal map $\sigma(u) = 1/au$
and define for $u \in \sigma(\Delta)$ the 
`almost translation', $\hat{f}(u) = \sigma \circ f \circ \sigma(u)$.
One verifies that the discrepancy, $\vartheta(u) = \hat{f}(u) - u - 1$,
satisfies a bound,
\beq |\vartheta(u)| < C |u|^{-\epsilon} , \ u \in \sigma(\Delta) \eeq
with $C<\infty$.

For $r>0$ small enough the disc $B=B(r,r)$, centered at
$z=r$ and of radius $r$, is contained
in $\Delta$ (since $\theta_0>\frac{\pi}{2}$) and
its closure is mapped strictly into its interior (union $\{0\}$).
For any $z\in \Delta$ there is a unique $n\in Z$ such that
$f^n(z) \in \bar{B} - f(\bar{B})$.
Choosing $r$ small enough we may ensure that the discrepancy
satisfies the bounds
\beq |\vartheta(u)| < 1/4\ , \ \ 
|\vartheta'(u)| < 1/4 \ , \ \ u \in \sigma(B) .\eeq
Using these bounds
the `almost' translation map $\hat{f}$
can be straightened out $K$-quasi-conformally
with $K<2$ on the  cylinder $\Phi=\sigma(\bar{B} - f(B))$
(\cite{Shishikura}, proof of Proposition A.2.1)
 and hence by uniformization
also conformally, i.e. $\psi(\hat{f}(u)) = \psi(u) + 1$,
where $\psi$ is univalent and holomorphic in $\Int \Phi$.
It extends homeomorphically, whence also holomorphically, to the whole of 
$\sigma(\Delta)$. Let $\Omega = \psi \circ \sigma (\Delta)$ 
and let $\phi : \Omega \rr \Delta$ be the inverse of $\psi \circ \sigma$.
One has $\phi (w+1) = \sigma (\hat{f} \sigma \phi^{-1}(w))=f(w)$
for $w\in\Omega$.
 From \cite{Shishikura} (Lemma A.2.4, case (i)) 
and our previous estimates it
follows that for $R_1>0$ big enough there are constants
$C_1,C_2>0$ such that if the disc $B(u_0,R)$, centered
at $u_0$ and of radius $R>R_1$, is contained in $\sigma (\Delta)$
then $|\psi'(u_0) - 1| \leq C_1/R + C_2 |u_0|^{-\epsilon}$.
As $\sigma(\Delta)$ contains the complex numbers of
the form $re^{i\theta}$ with $r>1/ar_0$ and $|\theta|< \theta_0$
it is verified that $\Omega$ contains 
angular (of any angle $\alpha$  strictly smaller than $\theta_0$) sectors 
$S_\alpha$ as described in the Lemma.
Finally, from the estimate on $\psi'$
we obtain that
 $\lim_{|u|\rr \infty} \psi'(u)=1$, whence
 $\lim_{|w|\rr \infty}a w^2 \phi'(w) = - 1$
where the convergence is uniform in any of the translated sectors.
 \Halmos \\
\mysubsect{The continuous spectrum.}
\label{cont-spect}
Let $\kappa(t) = \exp(-t)$. A continuous linear operator
   $M_\kappa    :  L^1(\RR_+)   \rr   L^1(\RR_+) $ is defined by~:
\begin{equation}
   M_\kappa\psi(t)  \equiv  \kappa(t) \psi(t) 
\end{equation}
Denote by $\Sp(M_\kappa)$ the spectrum of $M_\kappa$ and
for $\lambda \notin \Sp(M_\kappa)$,
let $R(\lambda,M_\kappa)=(\lambda-M_\kappa)^{-1}$ be the
associated resolvent operator. Define also
$\sigma_c = [0,1]$
   (the line segment in the complex plane). One has\\

\begin{Lemma}
\label{norm-lemma}
   {
   The spectrum of $M_\kappa$ is continuous and equals
    $\sigma_c = [0,1]$.
  \\}
\end{Lemma}

\noindent {proof :} As $\kappa(t)$ is continuous and bounded by 1,
 the operator $M_\kappa$ is a bounded linear operator. Whenever
$\lambda$ is not in the above mentioned line segment
  $\sigma_c$
 the function $(\lambda - e^{-t})^{-1}$ is continuous and bounded
 by $\dist(\lambda,\sigma_c)^{-1}$ thus providing an upper bound for the
 norm of the resolvent operator.
 By approximation one shows that it gives a lower bound as well.
 As it diverges for $\lambda$ approaching $\sigma_c$ we have 
 $\Sp(M_\kappa)=\sigma_c$. For $\lambda\in \CC$,
 $(\lambda-M_\kappa)u=0$ forces $u=0$ a.e. and hence,
 the point spectrum of $M_\kappa$ is empty.
 \Halmos\\

Let $m$ be a real number to be fixed below.
The (shifted) Laplace transform of a function $\psi$ in $L^1(\RR_+)$
is then given by~:
\begin{equation}
   {\cal L}_m \psi (w) = \lap \psi(t) e^{-(w-m)t}
\end{equation}

For $w \in H_m$ 
 the integral converges absolutely and uniformly and is
bounded by the $L^1$ norm of $\psi$. By standard arguments
${\cal L}_m \psi$ is analytic for $w \in H_m$ and has a continuous extension
to the boundary, its absolute value being bounded by the $L^1$ norm of $\psi$.
It is also clear that the map ${\cal L}_m:L^1(\RR_+) \rr {\cal H}(H_m)$
is injective. Hence we have
shown the following~:

\begin{Lemma}
 The map  ${\cal L}_m : L^1(\RR_+) \rr A(H_m)$ is bounded and injective.
 \Halmos
\label{a-lemma}
\end{Lemma}

 Denote by $X(H_m)$ the isometric
(with induced norm)  image of $L^1(\RR_+)$ in ${\cal H}(H_m)$
under the Laplace transform.\\

For functions $h \in {\cal H}(H_m)$ we define the
translation operator : $Sh(w) = h(w+1)$. 
Its restriction to $X(H_m)$ is conjugated to the multiplication
operator $M_\kappa$ (hence they have identical spectral properties)
 under the isometry given by the Laplace transform~:

\diagram     {L^1(\RR_+)} {M_\kappa} {L^1(\RR_+)} 
             {{\cal L}_m}                 {{\cal L}_m}
             {X(H_m)}    {S}    {X(H_m)}  
             {eq:L}

This follows since for $h = {\cal L}_m \psi$, $\psi \in L^1(\RR_+)$
and $w \in \bar{H}_m$ we have~:
\[ S{\cal L}_m\psi(w+1) = \lap e^{-(w-m)t-t}
     \psi(t) = {\cal L}_m M_\kappa \psi (w) .\]

Let  $K =
    \bigcup_{\omega \in \Xi}\Cl \; f_\omega (\Delta) \subset \Delta$
 be the closed union of the
 images of $\Delta$ under the
uniformly contracting family. We have

\begin{Lemma}
There is an open connected domain $N \subset \Omega$
and real numbers $m_1>m_2$
 such that $K \subset \phi (N)$,
$s(N) \subset N$ and  $\bar{H}_{m_1} \subset N \subset H_{m_2}$.
\label{N-lemma}
\end{Lemma}

\noindent proof : 
By Lemma \ref{conjug-lemma}  there is a sector
 $S_{\alpha=0} \supset  \bar{H}_{m_1}$ (for some $m_1 \in R$)
contained in $\Omega$.
Let $\gamma$ be a finite path
 connecting $\phi^{-1}K$ and $H_{m_1}$ in the open set $\Omega$. By compactness
there is an $n$ such that $s^n(\gamma \cup \phi^{-1}K) \in H_{m_1}$.
Let $\epsilon>0$ be such that the $\epsilon/(n+1-k)$ neighborhood
of $s^k(\gamma \cup \phi^{-1} K)$, $k \leq n$
 all are in $\Omega$. The union of these
and $H_{m_1}$ will satisfy our requirements with $m_2=m_1-n$. \Halmos\\

With the notation of the above Lemma
choose $m>m_1$ and let it be fixed in the following.
Let $\delta = m-m_1 >0$ and let $n$ be an integer such that
$m_2 + n > m$. The set $H_m$ shifted $n$ times to the left
covers the domain
$N$. Thus, we have 
\beq 
   s^{-n} H_m \supset N \supset H_{m-\delta} \supset H_m.
   \label{eq:inclu}
\eeq
Let $X(N)$ be the subset of functions in $X(H_m)$ which have an analytic
(necessarily unique) continuation to $N$ with a continuous
extension to the boundary. It is a Banach space under
the following norm~:
\begin{equation}
 \|h\|_{X(N)} \equiv \|r_{H_m}h\|_{X(H_m)} + \|h\|_{A(N)}  .\ 
\label{eq:norm}
\end{equation}
As $s(N) \subset N$ we may define the 
translation operator $T:X(N) \rr X(N)$ as $Th(w) = h(w+1)$.
The inclusions (\ref{eq:inclu})
 give rise (by analytic continuation and Lemma \ref{a-lemma})
 to bounded linear operators~:
\begin{equation}
    X(H_{m}) \stackrel{r_N S^n}{\longr} 
    X(N)     \stackrel{r_{H_m}}{\longr} X(H_m) .
\end{equation}
Denote $r=r_{H_m}$ and $q = r_N S^n$.
One verifies that
$q \circ S = T \circ q$,
$r \circ T = S \circ r$,
$r \circ q = S^n$,
$q \circ r = T^n$
and therefore  for $\lambda \neq 0$
\beq
  R(\lambda,S) =
   \frac{1}{\lambda} +
   \frac{S}{\lambda^2} + \cdots
   \frac{S^{n-1}}{\lambda^n} + \lambda^{-n} r \circ R(\lambda, T) \circ q
   \label{eq:resolvent}
\eeq
(and similarly with $S$ and $T$ interchanged). We conclude
that $\Sp(T) = \Sp(S) = \sigma_c$.\\

Remark : Life would be somewhat simpler if $K$ was contained
in a half plane $H_{m_2}$ which in turn was contained in $\Omega$.
However, this might not be the case (the
set $s^{-n}H_m$ in (\ref{eq:inclu}) might not be included in $\Omega$)
in which case 
the above  distinction between the two translation
operators $S$ and $T$ proves necessary.\\

The conjugating map $\phi$ (Lemma \ref{conjug-lemma})
 is univalent and $\Omega$ is simply connected.
The image of $\Omega$ under $\phi'$ is thus contractible in $\CC^*=\CC-\{0\}$.
Hence $\log(-\phi'(w))$ is uniquely determined when specifying
its asymptotic behavior in $H_m$,
 $\lim_{w \rr \infty} \log (- a w^2 \phi'(w)) = 0$.
For $\beta \in \CC$ a complex parameter
we set $(-\phi')^\beta = \exp(\beta \log(-\phi'))$.
The same arguments apply to $f:\Delta \rr \Delta$ and
$(f')^\beta$ is uniquely determined when specifying 
$\log(f'(0))=0$.
Given any open set $\Upsilon \subset \Omega$ we may define the
following isomorphism $J : {\cal H}(\phi \Upsilon) \rr {\cal H}(\Upsilon)$~:
\beq
   J v  (w) \equiv v \circ \phi(w) \; (-\phi'(w))^\beta .
\eeq

Let $U = \phi N \subset \Delta$ and
denote by $X(U) \subset {\cal H}(U)$
 the isometric (with induced norm) image of $X(N)$ under
$J^{-1}$.
The translation operator $T:X(N)\rr X(N)$ is then conjugated to
the Perron-Frobenius type operator, ${\cal M}_0 : X(U) \rr X(U)$~:
\begin{equation}
   {\cal M}_0 v (z) = v \circ f (z) \; (f'(z))^\beta .
\end{equation}
This is seen from $ \phi'\circ s = f' \circ \phi \; \phi' $ and hence
\beq J{\cal M}_0 v =  (v \circ f \cdot (f')^\beta)
               \circ \phi \cdot (-\phi')^\beta
          = (v \circ \phi \cdot (-\phi')^\beta) \circ s = T J v .\eeq
\diagram
             {X(N)} {T} {X(N)}
             {J^{-1}}{J^{-1}}
             {X(U)} {{\cal M}_0} {X(U)}
             {eq:J}
In particular, it follows that the spectrum
of ${\cal M}_0 : X(U) \rr X(U)$
is the line-segment $\sigma_c$. One verifies that
the point spectrum of ${\cal M}_0$ is void.

\noindent Remark :
The function spaces $X(\cdot)$ are in some sense quite abstract
as they are given through isometric
injections into Fr\'echet spaces. In contrast, the spaces
$A(\cdot)$ are very easy to handle.  
The following two Lemmas provide the crucial relationship
between these function spaces.\\

\begin{Lemma}
 \label{injection-lemma}
 { For Re$\;\beta > \half$ the injection map 
   $A(U) \stackrel{j}{\hookrightarrow} X(U)$ is continuous.}
\end{Lemma}

\noindent {proof :} Let $\tau =$ Re$\; \beta$.
 For $v \in A(U)$ both $v\circ \phi$ and $\phi'$ belong to $A(N)$,
whence $\|Jv\|_{A(N)} \leq \|v\|_{A(U)}
      \|\phi'\|_{A(N)}^{\tau}$.
Since also $w^2 \phi' \in A(N)$ and $H_{m-\delta} \subset N$
we may retrieve (an $L^1$ representative of)
the inverse Laplace transform of $Jv$.
Using a change of coordinates 
$u \rr u-\delta$  and analyticity we obtain~: 
\beq \psi(t) = \int_{\partial H_{m-\delta}} \frac{du}{2\pi i} e^{t(u-m)} Jv(u)
  = \int_{\partial H_m} \frac{du}{2\pi i}
         e^{-t\delta+ t(u-m)} Jv(u-\delta)\eeq
and since $(1+|u|^2)|\phi'(u)| \leq
           \|\phi'\|_{A(N)} + \|w^2\phi'\|_{A(N)} \equiv  c < \infty$,
for $u \in N$,
we have that
\beq |\psi(t)| \leq e^{-t \delta} \|v\|_{A(U)}
       \int_{-\infty}^{\infty}
       \frac{dx}{2\pi} \frac{c^{\tau}}{(1+x^2)^\tau}
 = C(\tau) e^{-t\delta} \|v\|_{A(U)} 
\eeq
where $C(\tau) < \infty$ since $\tau > \half$.
Integration yields
 $\|\psi\|_{L^1(\RR_+)} \leq C(\tau) \frac{1}{\delta} \|v\|_{A(U)}$.
As $N$ is connected, $Jv$ is determined uniquely through its
inverse Laplace transform $\psi_t$ and we obtain~:
\beq
\|Jv\|_{X(N)} = \|Jv\|_{A(N)} + \|r_{H_m} Jv\|_{X(H_m)}
      \leq \|v\|_{A(U)} (\|\phi'\|^\tau_{A(N)} + \frac{C(\tau)}{\delta}) .
\eeq
Composing with the isometry $J^{-1}:X(N)\rr X(U)$
yields the Lemma.
\Halmos\\

For the other direction we have the following

\begin{Lemma}
\label{restriction-lemma}
For any compact set $K \subset U$ the restriction map
$r_K : X(U) \rr A(K)$ is continuous.
\end{Lemma}

\noindent proof : 
Set $K = \phi Q$. For $v \in X(U)$, $h =Jv \in X(N)$
we have $p_Q(h) \leq p_N(h) \leq \|h\|_{X(N)} = \|v\|_{X(U)}.$
Hence
\beq
\|r_K v\|_{A(K)} =
 p_K(v) \leq
 p_{K}(J^{-1}h) \leq p_Q(h) p_{Q}((-\phi')^{-\beta}) 
  \leq \|v\|_{X(U)} C(\beta)
\eeq
where $C(\beta) < \infty$ since $Q$ is compact.
\Halmos\\

\mysubsect{Nuclear theory and regularized Fredholm determinants.}

\begin{Lemma}
\label{nuc-lemma}
For Re$\;\beta > \half$ and  $\lambda \in \CC-\sigma_c$ the operator
${\cal M}_1 R(\lambda,{\cal M}_0) : X(U) \rr X(U)$ is nuclear
of order zero.
\end{Lemma}

\noindent proof :
As above let $K$ be the closure of union of the images of $\Delta$
under the uniformly contracting family and
choose $K'$ compact such that
$K \subset  \Int \; K' \subset K' \subset U$.
 The injection
$A(K') \hookrightarrow {\cal H}(\Int K')$ is a continuous
map and the further restriction $r_K : {\cal H}(\Int K')
 \rr A(K)$ is a bounded linear map from the nuclear space
${\cal H}(\Int K')$ to the Banach space $A(K)$. It follows
that $r_K : A(K') \rr A(K)$ is nuclear of order zero
\cite{Grot55,Ruelle76}.

 From the integrability
condition on the weights 
the linear operator
(from equation \ref{eq-PFlast}),
$\tilde{\cal M}_1:A(K) \rr A(U)$ is bounded
and hence by Lemma \ref{injection-lemma} and
\ref{restriction-lemma} so is 
${\cal M}_1=j \tilde{\cal M}_1 r_K :X(U) \rr X(U)$.
For $\lambda \notin \sigma_c$ the resolvent operator
$R(\lambda,{\cal M}_0):X(U) \rr X(U)$ is bounded and
so is $r_{K'}R(\lambda,{\cal M}_0):X(U) \rr A(K')$
by Lemma \ref{restriction-lemma}. Composing with the
further (nuclear) restriction $r_K : A(K') \rr A(K)$,
the bounded operator $\tilde{\cal M}_1 : A(K) \rr A(U)$ and
finally the continuous injection  (Lemma \ref{injection-lemma}),
$j : A(U) \hookrightarrow X(U)$ we conclude that
\beq
{\cal M}_1 R(\lambda,{\cal M}_0) =
   j \tilde{\cal M}_1 r_K r_{K'} R(\lambda,{\cal M}_0) : X(U) \rr X(U)
\eeq
is nuclear of order zero.
\Halmos\\

For $\lambda_0 \in \CC-\sigma_c$, Re$\;\beta_0 > \half$ the
bounds in lemma \ref{injection-lemma},
\ref{restriction-lemma} and \ref{nuc-lemma}
can be made uniform in a small complex neighborhood of $\lambda_0$ and
$\beta_0$ (more precisely, one should consider the operator
$\tilde{\cal M}_1 r_K R(\lambda,{\cal M}_0) j : A(U) \rr A(U)$
for which the function space is kept fixed).
 It follows that the
operator ${\cal M}_1 R(\lambda,{\cal M}_0):X(U) \rr X(U)$ is
a holomorphic family of nuclear operators 
in $\lambda \in \CC-\sigma_c$, Re$\;\beta > \half$ and by
\cite{Grot55,Grot56} it has a Fredholm determinant
\beq
   \hat{d}(z,\lambda,\beta) = \det (1 - z {\cal M}_1 R(\lambda,{\cal M}_0)) 
   \label{eq:fredh}
\eeq
which is entire in $z$ and holomorphic in 
 $\lambda \in \CC-\sigma_c$, Re$\;\beta > \half$.
It corresponds to the Fredholm determinant associated
with the so-called induced family of contractions.
Setting $z=1$ we obtain the regularized Fredholm determinant
of ${\cal M}={\cal M}_0 + {\cal M}_1$~:
\beq
   {d}_\beta(\lambda) = \det (1 -  {\cal M}_1 R(\lambda,{\cal M}_0)) 
\eeq
 In order to relate its zeroes to the
eigenvalues of the operator ${\cal M}={\cal M}_0+{\cal M}_1$ we need
the following~:

\begin{Lemma}
 \label{nuclear-lemma}
 Let $A : X \rr X$ be a bounded linear operator on a Banach space $X$
and let $N : X \rr X$ be nuclear of order zero. Assume that
$E = \CC - \Sp(A)$ is connected. Then the
part of the spectrum of $A+N$ which intersects
$E$ consists of isolated eigenvalues
of finite multiplicity only which can not accumulate in $E$.
 The Fredholm determinant
 \[ d(u) = \det(1 - N (u - A)^{-1}) \]
is analytic in $u \in E$.
Furthermore in this domain, the zero-set of $d(u)$ counted with
order is the same as
 the eigenvalues of $A+N$ counted with multiplicity.
\end{Lemma}

\noindent {proof :}
For finite matrices the result is trivial. The
problem arises since the determinant of $(1 - u^{-1} A)$ is not defined.
Note first, that for $\lambda \in E$, the resolvent
$R(\lambda) = (\lambda-A)^{-1}$ is a (bounded)
 operator-valued analytic function of $\lambda$.
The expression
\[ (\lambda - A-N)^{-1} = R(\lambda) (1 - N R(\lambda))^{-1} \]
is valid if $1$ is not in the spectrum of
$N R(\lambda)$. The operator $N R(\lambda)$ is a nuclear operator
 valued (being  a nuclear operator times a bounded operator)
analytic function of $\lambda \in E$. Its spectrum is discrete
and for $\lambda \rr \infty$ all eigenvalues go to zero. In particular
as $E$ is connected the value $1$ can be an eigenvalue of $N E(\lambda)$
only for a discrete set of $\lambda$ values in $E$.\\

From the above it follows \cite{Grot55,Grot56}
 that the Fredholm determinant $d(u)$ 
is analytic in $u \in E$. A root of $d$, $d(\lambda)=0$,
corresponds to  an eigenvalue 1 of $N R(\lambda)$. Let $\lambda \in E$
be a (necessarily isolated) zero of $d$ of order $p$ and let ${\cal C}$ be the
boundary of a small closed disc
$E$ which intersects 
the spectrum of $A+N$ only in $\lambda$.
Through analytic continuation one obtains from
 $ d(u) = \exp ( \tr (\log (1 - N R(u)))) $  the expression
\[ \frac{d}{du} \log d(u)   =   \tr [(1-NR(u))^{-1} NR(u)^2] . \]
 Integrating along the contour ${\cal C}$ we get
\begin{eqnarray}
 p & = &
    \oint_{\cal C} \frac{du}{2\pi i} \frac{d}{du} \log d(u) 
           \nonumber \\
    & = &  \tr \oint_{\cal C} \frac{du}{2\pi i} [(1-NR(u))^{-1} NR(u)^2]
           \label{eq:nuc} \\
    &=&   \tr \oint_{\cal C} \frac{du}{2\pi i} [(1-NR(u))^{-1} R(u)]
           \label{eq:id} \\
    &=&   \tr \oint_{\cal C} \frac{du}{2\pi i} [(u - A - N)^{-1}]
           \label{eq:fin} \\
    &=&   \tr P_\lambda
           \nonumber
\end{eqnarray}
where $P_\lambda$ is the projection onto the
generalized eigenspace of $A+N$ corresponding to the eigenvector $\lambda$.
The validity of the above calculations is justified by
the  nuclearity of $N$ (eq. \ref{eq:nuc}), the identity
\[ (1-NR(u))^{-1} NR(u)^2 = -R(u) + (1-NR(u))^{-1} R(u)  \]
and the analyticity of $R(u)$
(thus yielding a vanishing contribution)
 inside the contour (eq. \ref{eq:id}) and
the finite dimensionality of the resulting projection operator 
(eq. \ref{eq:fin}).
\Halmos\\
\label{nuc-section}

For $|\lambda| > \|{\cal M}_0\| + \|{\cal M}_1\|$ we have the uniformly
convergent series~:
\beq
 \log (1 - \Mone (\lambda - \Mzer)^{-1})
      = \Mone \sum_{m >  0} \lambda^{-m} A_m
\eeq
where $A_m$ is a polynomial in ${\cal M}_0$ and ${\cal M}_1$
 for which 
$ \|A_m\| < (\|{\cal M}_0\| + \|{\cal M}_1\|)^{m-1}$.
We note that the right hand side is nuclear of order zero and 
by the formal calculation
$(1 - \Mone(\lambda - \Mzer)^{-1}) =
 (1 - \lambda^{-1}(\Mzer + \Mone))(1-\lambda^{-1}\Mzer)^{-1}$
we derive the formula~:
\beq
 \tr \;
 \log (1 - \Mone (\lambda - \Mzer)^{-1})
      =  \sum_{m > 0} \frac{\lambda^{-m}}{m} 
        \tr \; ((\Mzer + \Mone)^m - \Mzer^m) .
\eeq
Defining $d_m = 
        \tr \; ((\Mzer + \Mone)^m - \Mzer^m)$ we conclude
that the Fredholm determinant~:
\beq
  d(\lambda) = \det
  (1 - \Mone (\lambda - \Mzer)^{-1})
      = \exp - \sum_{m > 0} \frac{\lambda^{-m}}{m} d_m
\eeq
is holomorphic for $|\lambda| > 
\|{\cal M}_0\| + \|{\cal M}_1\|$. By 
Lemma \ref{nuc-lemma}
it has a unique analytic extension to $\CC-\sigma_c$.
Inserting (\ref{eq-PFfirst}-\ref{eq-PFlast})
and using the uniform contraction
 and the uniform convergence of the integrals 
(hence uniform nuclearity)
we get
for the trace~:
\beq
  d_m = \tr \int_{\Xi^*_m} \mu^*(\bo) \; {\cal M}^m_\bo
   = \int_{\Xi^*_m} \mu^*(\bo) \; \tr \; {\cal M}^m_\bo ,
\eeq
where the multiple-integral extends over all
$\bar{\omega} = (\omega_1,...,\omega_m) \in \Xi^*_m
= \Xi^m - \{ (0,...,0) \}$ and
${\cal M}^m_{\bar{\omega}}=
  {\cal M}_{\omega_m}
   \cdots
  {\cal M}_{\omega_1} : X(U) \rr X(U)$ is given by
${\cal M}^m_{\bar{\omega}} v(z) = g_{\bar{\omega}}(z) \;
   v ( f_{\bar{\omega}}(z))$.
As $f_{\bar{\omega}} : \Cl \; U \rr \Int  \;U$ standard arguments
show that this map has a unique fixed point $x(\bar{\omega}) \in U$
where its multiplier is strictly smaller than one in modulus.
Furthermore, ${\cal M}^m_{\bar{\omega}}$
is a nuclear operator on $X(U)$ and its trace does not
change if we replace $X(U)$ by $A(D)$
where the domain $D$ is
such that for some $k>0$, $\Cl \; f^k_{\bar{\omega}} U \subset \Cl \; 
 f_{\bar{\omega}} D \subset D \subset U$.
Choosing for $D$ a small disc centered at $x(\bar{\omega})$
we obtain by standard Fredholm theory and residue calculus,
\beq
 \tr  {\cal M}^m_{\bar{\omega}}
 = \oint_{\partial D} \frac{dz}{2\pi i}
       \frac{g_{\bar{\omega}}}{z-f_{\bar{\omega}}(z)} =
   \frac{g_{\bar{\omega}}(x(\bar{\omega}))}
        {1 - f'_{\bar{\omega}}(x(\bo))} ,
\label{eq-traces}
\eeq
from which the formula (\ref{eq:fredcoef}) for the
traces follows.
\mysubsect{Crossing the continuous spectrum.}

 With $\theta_0$
as in property (ii)
and for $\alpha \in (\frac{\pi}{2},\theta_0)$
 let ${N}_\alpha$ be the union of
$N$ and the  sector $S_\alpha$
(equation \ref{eq:sector}, Lemma \ref{conjug-lemma}). A 
Banach space $X({N}_\alpha)$ and a norm $\|\cdot\|_{X({N}_\alpha)}$
is defined as in equation (\ref{eq:norm}).
Let ${U}_\alpha=\phi {N}_\alpha$ and denote again by
$X({U}_\alpha)$ the isometric image of $X({N}_\alpha)$ under
$J^{-1}$.
As $\phi'$ and $w^2 \phi'$ are uniformly bounded in ${N}_\alpha$,
a repetition of the proof of Lemma \ref{injection-lemma}
shows that
\begin{Lemma} For Re$\; \beta>\half$ 
 the injection map
   $A({U}_\alpha)
     \stackrel{{j}_\alpha}{\hookrightarrow} X({U}_\alpha)$ is continuous.
   \Halmos
\label{ext-lemma}
\end{Lemma}

For $\nu \in (\frac{\pi}{2}-\alpha,
\alpha-\frac{\pi}{2})$ set $\kappa = \tan \nu$ and define
$Y^\nu$ as the set of complex numbers $\gamma$ where
either Re$\;\gamma <0$ and Im$\; \gamma \in [0,2\pi]$ 
 or Re$\;\gamma \geq 0$ and
 Im$\gamma - \kappa  \mbox{Re} \gamma \in (0,2\pi)$.
Under the exponential map
$\lambda = \exp(-\gamma)$,
 $2\pi i$ separated points are identified and
 the strip $Y^0$ maps
to the sliced complex plane $\CC-\sigma_c$
and $Y^{\nu\neq 0}$ maps to $\CC-\{\mbox{a logarithmic spiral}\}$.

We let $Y_{\alpha}$ denote the union of all $Y^\nu$, 
$|\nu| < \alpha-\frac{\pi}{2}$ (which then maps to a multi-sheeted
Riemann surface under the exponential map).
Finally, define
the domain $T_\alpha = \{r e^{i \theta} : r>0, \theta \in
 (\frac{\pi}{2}-\alpha,\alpha-\frac{\pi}{2})\}$.

\begin{Lemma}
Let $K' \subset U$ be a compact set.
The operator $Q(\gamma) : A({U}_\alpha) \rr A(K')$ given by~:
\beq Q(\gamma)   = r_{K'} \circ R(e^{-\gamma},{\cal M}_0)  \circ r_U 
         \circ {j}_\alpha .
\eeq
has a unique holomorphic extension to the domain $Y_\alpha$.
\label{Q-lemma}
\end{Lemma}

\noindent [Remark :
Our previous analysis shows already that $Q(\gamma)$ 
is a bounded operator for $\gamma \in Y^0$, i.e. for
$\lambda =e^{-\gamma} \notin \sigma_c$. What is new is that
the borders $\RR_+$ and $\RR_+ + 2 \pi i$
may be crossed (within some neighborhood)
 without encountering any singularities
from the continuous spectrum of ${\cal M}_0$.]\\

\noindent proof : 
 We may assume
that $K' \in H_m$ or else use the technique leading
to equation (\ref{eq:resolvent}) to achieve this situation.
Assume first that $\lambda = e^{-\gamma} \notin \sigma_c$.
For a function $v \in A(U_\alpha)$ we retrieve
the inverse Laplace transform of $h = Jv$
 as in Lemma \ref{injection-lemma}
by integrating along the boundary of $H_{m-\delta}$.
Since
 $\phi'$, $w^2 \phi'$ and $j_\alpha v$ are uniformly bounded in ${N}_\alpha$
this integration may be pushed to a contour
${\cal C}_\alpha$ which is formed by taking the boundary of
the union of $H_{m-\delta}$ and the
angular sector $S_\alpha$ of Lemma \ref{conjug-lemma}
(note that Re$(w-u) \geq \delta > 0$ for $w\in H_m$ and
 $u \in {\cal C}_\alpha$).
Let $k_\gamma(w,t) = e^{-wt}/(e^{-\gamma} - e^{-t})$
be the kernel (cf. the proof of Lemma \ref{norm-lemma})
 for the resolvent operator $R(e^{-\gamma},{\cal M}_0)$.
For $\gamma\in Y^0$ and $w \in H_m$ the action of $Q(\gamma)$
on $v\in A(U_\alpha)$ is given
by~:
\beq 
   [J Q(\gamma) v] (w)  = 
     \int_0^\infty dt \;
     \int_{{\cal C}_\alpha} \frac{du}{2\pi i}
 k_\gamma(w-u,t)
        Jv(u) .
 \label{eq:JQ}
\eeq
[Remark : 
The point in pushing the $u$-integration to the contour ${\cal C}_\alpha$
is that
when $w\in H_m$ and $C>0$ is arbitrary
 the exponential term in the kernel, $\exp(-(w-u)t)$,
remains uniformly bounded for all (now complex) values of $t \in T_\alpha$
with Im$\; t \in (-C,C)$ 
(since
  $u$ asymptotically behaves like $re^{\pm i \alpha}$, $r \rr \infty$).
 This provides
us with some flexibility in deforming the integration path for the
$t$-variable.]\\
For $\epsilon \in (0,\frac{\pi}{4})$,
 $W^\epsilon_\gamma = \bigcup_{k \in Z}
    B(\gamma+2 \pi i k, \epsilon)$ will denote the union of
disjoint $\epsilon$-balls centered around the
singularities of $t \mapsto k_\gamma(w-u,t)$
(the zeroes of $e^{-\gamma}-e^{-t}$).

Let ${\cal P}^\epsilon_\gamma$ be the set of smooth
(denoted admissible) paths in $T_\alpha - W^\epsilon_\gamma$ which 
starts at the origin and tends to $+\infty$ asymptotically
parallel to the real axis.
 For any such path $\Gamma \in 
{\cal P}^\epsilon_\gamma$, the kernel
$t \mapsto k_\gamma(w-u,t)$ stays uniformly bounded
(avoiding the singularities) for
$t \in \Gamma$, $w \in H_{m}$, $u \in {\cal C}_\alpha$
and furthermore, it tends to zero exponentially fast as Re$\;t \rr \infty$.
For any $\gamma \in Y_\alpha$ we can find
a suitable $\epsilon>0$ and connect $\gamma$ by a path to
some fixed  $\gamma_0 \in Y^0$ and by following this connection
in the opposite direction we deform 
at the same time smoothly 
the integration 
path $\Gamma(\gamma_0) = \RR_+\in {\cal P}^\epsilon_{\gamma_0}$
to an integration  path $\Gamma(\gamma) \in {\cal P}^\epsilon_{\gamma}$ 
through admissible paths only and in the process obtain an
 analytic continuation $Q(\gamma)$  of $Q(\gamma_0)$. 
As $Y_\alpha$ is simply connected this extension is unique.\Halmos\\

\begin{Lemma}
\label{extension-lemma}
For Re$\;\beta > \frac{1}{2}$
the regularized Fredholm determinant 
extends holomorphically through the map 
$\gamma \mapsto d(\lambda = e^{-\gamma})$
to the domain $Y_\alpha$ described above.
 \end{Lemma}

\noindent proof : Noting that ${\cal M}_1$ extends to a continuous
operator ${\cal M}_1 : A(K) \rr A(U_\alpha)$ the sequence,
\beq
   A(U_\alpha) \stackrel{Q(\gamma)}{\longr}
   A(K') \stackrel{r_K}{\longr}
   A(K) \stackrel{{\cal M}_1}{\longr}
   A(U_\alpha) ,
\eeq
defines a nuclear operator valued holomorphic function in $\gamma\in Y_\alpha$
(cf. the proof of \ref{nuc-lemma}). One verifies that its
Fredholm determinant on the strip $\gamma \in Y^0$ coincides
with $d(\lambda=e^{-\gamma}, z,\beta)$ from  equation (\ref{eq:fredh}).
\Halmos\\

For $\nu \in (\frac{\pi}{2}-\alpha,
\alpha-\frac{\pi}{2})$ set $\eta=e^{i\nu}$
and consider for $\psi \in L^1(\RR_+)$
 its rotated and shifted Laplace transform~:
\beq
    {\cal L}_{m,\eta} \psi (w) = \lap \psi(t) e^{-(w-m)\eta t} .
\eeq
It is holomorphic in 
\beq
    H_{m,\eta} = \{ w : \mbox{Re} ((w-m)\eta) > 0 \}
\eeq
and has a continuous extension to the boundary.
The translation operator $S_\eta$ on the  space $X(H_{m,\eta})$
is now conjugated to a multiplication
with $\exp(-\eta t)$ on $L^1(\RR_+)$ and
we deduce that the continuous
spectrum now becomes a logarithmic spiral~:

\beq
  \Sp(S_\eta) = \Cl \{e^{-\eta t} : t \in \RR_+\} .
\eeq

Replacing $H_m$ by $H_{m,\eta}$,
our previous analysis carries over and provides us with
a rotated domain $N_\eta$ (Lemma \ref{N-lemma},
possibly with other choices of constants, $m_1$ and $m_2$) and a
Banach space $X(U_\eta)$ on which ${\cal M}_0$ and
${\cal M}_1$ act as bounded linear operators such that
$\Sp({\cal M}_0) = \Sp(S_\eta)$ and ${\cal M}_1$ is nuclear as before.
Then ${\cal M}_1 R(\lambda,{\cal M}_0)$ has a Fredholm
determinant $d_\eta(\lambda)$ which is analytic in
$\CC-\Sp(S_\eta)$ where its zero-set is in one-to-one correspondence
with the eigenvalues of ${\cal M}={\cal M}_0 + {\cal M}_1 
\in L(X(U_\eta))$. As the formula (\ref{eq-traces})
for the traces remains the same, the analytic continuation
of $\gamma \rr  d_\eta(e^{-\gamma})$, $\gamma \in Y_\alpha$ is
independent of $\eta$. We use this to deduce the following

\begin{Lemma}
\label{spectrum-lemma}
For $\lambda_0\in \Sp(S_\eta)-\{0,1\}$
 let $q$ be the dimension of the generalized
eigenspace $\bigcup_{k>0} \mbox{\rm ker} (\lambda_0-{\cal M})^k$
(in $X(U_\eta)$).
Then the Fredholm determinant $d$ when analytically
 continued from either side of $\Sp(S_\eta)-\{0,1\}$ will have a
zero of at least  the same order $q$. In particular,
the point spectrum of ${\cal M}$ (in $X(U_\eta)$) can not accumulate
on points other than $\{0\}$ and $\{1\}$.
\end{Lemma}

\noindent proof : 
Write $\lambda_0 = e^{-\eta \gamma}$ with $\gamma \in (0,\infty)$ and
let $v \in X(U)$ with $v \in \mbox{\rm ker} (\lambda_0-{\cal M})^k$,
for some $k \geq 1$. Put $v_n = (\lambda_0-{\cal M})^n v \in X(U)$
for $n\geq 0$ so that $v_k=0$ and
\beq
 (\lambda_0- {\cal M}_0) v_n = v_{n+1} + {\cal M}_1 v_n .
\eeq
Choose $\epsilon>0$ small enough so that
$H_{m-\delta,\eta+\tau} \subset N_\alpha$
 for some $\delta>0$
and all $|\tau|<\epsilon$
 and such that
the function $\lambda_0 - e^{-\eta t}$ 
is non zero for $t\in \Cl\; T_\delta$ except at $t=\gamma$.
The operator ${\cal M}_1$ maps $A(K)$ and therefore also $X(U)$ into 
$X(U_\alpha)$ 
and like in Lemma \ref{injection-lemma} and \ref{ext-lemma},
${\cal M}_1 v_n$ has
an inverse Laplace transform
(an $L^1(\RR_+)$ representative) $\hat{\rho}_n(t)$ which 
 has an analytic continuation to $t \in T_{\delta}$
and decays exponentially as $|t|\rr \infty$.
Let $v_n = {\cal L}_{m,\eta}\hat{v}_n(t)$. Then
\beq
 \int_0^\infty dt \; e^{-(w-m)\eta t} ((\lambda_0-e^{-\eta t})\hat{v}_n(t) - 
             \hat{\rho}_n(t) - \hat{v}_{n+1}(t)) = 0 \ \mbox{(a.e.)}
\eeq
implies that $((\lambda_0-e^{-\eta t})\hat{v}_n(t) - \hat{\rho}_n(t)) 
 - \hat{v}_{n+1}(t)
 = 0$ a.e. By induction (starting from $v_k$) we see that 
all $\hat{v}_n$ extends analytically to $T_\delta$, possibly
with pole singularities at $t=\gamma$.
If however,  $\hat{\rho}_n(\gamma)+\hat{v}_{n+1}(\gamma)$ did not vanish, then
$\hat{v}_n$ would not be in $L^1(\RR_+)$. Hence,
the apparent singularity
 at $t=\gamma$ is removable and
 $\hat{v}_n(t)$
has an analytic continuation to $T_\delta$.
As $\hat{v}_n(t)$ also decays exponentially for $|t|\rr\infty$
it follows that $v_n$ is a generalized eigenvector also
for the operator ${\cal M}$ acting on $X(U_{\eta+\tau})$,
$|\tau|<\epsilon$. By Lemma \ref{nuclear-lemma}
the Fredholm determinant
$d_{\eta+\tau}(\lambda)$ has
for $0<|\tau|<\epsilon$
 a zero of order at least $q$
at $\lambda=e^{-\eta\gamma}$ and since independent
of $\eta$ the same is true for $d_\eta(\lambda)$ when analytically
continued from either side (corresponding to positive and negative
values of $\tau$). By analyticity of $d_\eta(\lambda)$ in a neighborhood
of $\lambda=\lambda_0$ the claim about
accumulation points for the point spectrum of ${\cal M}$ now follows.
\Halmos
\mysubsect{Proof of the Theorem and Corollaries. The parabolic case.}

The operator ${\cal M}_0$ has a continuous spectrum $\sigma_c=[0,1]$.
As ${\cal M}= {\cal M}_0 + {\cal M}_1$ is a perturbation
with a nuclear (in particular, compact) operator,
we obtain part (a) of the Theorem except that we have only
shown it for the part of $\sigma_p$ which is contained in $\CC-[0,1]$.
Lemma \ref{nuc-lemma}
provides the holomorphic continuation of the
regularized Fredholm determinant obtained in section
\ref{nuc-section}
 and Lemma \ref{nuclear-lemma}
 identifies its
zero set with the eigenvalues of ${\cal M}$ thus proving
part (c). 
By Lemma \ref{extension-lemma}, $d(\lambda)$ has analytic
continuations from both sides to an open neighborhood
of $\sigma_c-\{0,1\}$, proving part (d) and 
by Lemma \ref{spectrum-lemma} we see that eigenvalues
in $\CC-\sigma_c$  of ${\cal M}$  can accumulate only
on the points $0$ and $1$, proving part (b) and thus finishing
the proof of the Theorem in the parabolic case with
weight $g=(f')^\beta$. \Halmos \\

Composing with $z \mapsto \lambda(z) = \frac{1}{4z}(1+z)^2$
we obtain
\beq
 d(\lambda(z)) = \det (1 - {\cal M}_1 ( \frac{1}{4z}(1+z)^2
          - {\cal M}_0)^{-1} ) =
 \det (1 - 4 z {\cal M}_1 ( (1 + z)^2 -
           4 z {\cal M}_0)^{-1} ) 
 ,
\eeq
holomorphic for $\lambda(z) \in \CC-\sigma_c$ and hence
for $0 < |z|< 1$. This extends holomorphically to the origin,
proving Corollary \ref{conformal-corollary}.\\

Lemma \ref{extension-lemma} shows that $Q(\gamma)$ and hence
$d(\lambda=e^\gamma)$ has an analytic continuation to
$\gamma \in Y_\alpha$ for any $\alpha<\pi$ and hence to $\gamma \in Y_\pi$.
Taking the exponential of $Y_\pi$ yields the Riemann surface
described in Corollary \ref{extension-corollary}.\\

Under the conditions of Corollary \ref{zeta-corollary}
we may replace the weights $g_\omega$ by $g_\omega (f'_\omega)^\nu$
for Re$\;\nu\geq 0$. Making this $\nu$ dependence explicit
the operator ${\cal M}(\nu)$ gives rise to a holomorphic
family of regularized Fredholm determinants for Re$\;\nu>0$.
Since one has
\beq
 \zeta_m (\nu=0) = (g(0))^m + d_m (\nu=0) - d_m(\nu=1)
\eeq
where $g(0)=1$ we get that
\beq
 \zeta(z) = \frac{1}{\log(1-z)} \frac{d(1/z,\nu=1)}{d(1/z,\nu=0)}
\eeq
from which the Corollary \ref{zeta-corollary} follows.\\

\mysubsect{Reduction to the parabolic case.}
\label{reduction}
As $\Delta$ is simply connected in $\CC^* = \CC - \{0\}$ the conformal
transformation $u = u(z) = z^{p}$ of $\Delta$ is well-defined, (for $p>1$ 
possibly with a multi-sheeted image, which, however, is unimportant for
the arguments).
The map $f$ then conjugates to~:

\begin{equation}
    u \circ f \circ u^{-1} =
      u - pa u^2 + {\cal O}(|u|^{2+\epsilon})
\end{equation}
and the weight becomes~:
\begin{equation}
   g \circ u^{-1} = 1 - b u + {\cal O}(|u|^{1+\epsilon})
\end{equation}
(and similar transformations for $\{f_\omega,g_\omega\}_{\omega\in\Xi}$).

In the following we shall consider these conjugated
maps (denoted by the same symbols).\\

As ${g} \neq 0$ on the invariant set $\Sigma^*$ it does not vanish in
an open neighborhood of $\Sigma^*$ in $\Delta \cup \{0\}$.
 Possibly by replacing
$\Delta$ by a smaller set obtained by iterations under the extended
family we may assume that $g\neq 0$ on $\Delta$.\\

 We set (cf. section \ref{cont-spect})
\begin{equation}
  (f'(u))^\beta = \exp (\beta \log (f'(u)) , \  u \in \Delta
\end{equation}
with $\log(f'(0))=0$
and for $u \in U$ (defined as in section \ref{cont-spect}) we have
\begin{equation}
 (f'(u))^\beta = 1 - 2 a  p  \beta  u + 
            {\cal O}(|u|^{1+\epsilon}) .
\end{equation}
Choosing now $\beta = \frac{b}{2 p a}$, we may write
\begin{equation}
  g(u) = (f'(u))^\beta \psi(u)
\eeq
where $\psi(u) = 1 + {\cal O}(|u|^{1+\epsilon}) \in A(U)$ 
does not vanish. From the asymptotic behavior of $f$,
$k f^k(u)$ is uniformly bounded for $u \in\Cl(U)$ and $k \geq 0$.
It follows that the product $\prod_{k \geq 0} \psi(f^k(u))$
converges uniformly. The limit $\psi^* \in A(U)$ does not vanish
and hence we have also $1/\psi^* \in A(U)$.\\

The calculation
\beq
 {\cal M}_0 (\psi^* v) = (\psi^*\circ f) \; (v \circ f) \; 
      (f')^\beta \; \psi = \psi^* \; (v \circ f) \; (f')^\beta
\eeq
shows that ${\cal M}_0$ is conjugated 
to an operator where the weight has been replaced by $(f')^\beta$
and the norm on the function space is obtained by scaling
with $\psi^*$.
If we assume that $\mbox{Re}(\beta) > \half$ then since
$\psi^*,1/\psi^* \in A(U)$ the traces
of ${\cal M}^m_{\bar{\omega}}$, $\bar{\omega} \in \Xi^*_m$ remains
the same. 
Noting that the condition Re$(\beta) > \half$
 translates into Re$\;b > pa$
and $\theta_0$ maps into $p\theta_0$ (cf. condition (ii)
and Corollary \ref{extension-corollary}),
the Theorem and its Corollaries follow.

\renewcommand{\baselinestretch} {1}


\begin{thebibliography}{99}
{\small


% \bibitem{Carleson} L. Carleson and T.W. Gamelin,
        % {\em Complex Dynamics} (Springer-Verlag, 1993).

\bibitem{Grot55} A. Grothendieck,
        {\em Produits tensoriels topologiques et espaces nucl\'eaires},
        Memoirs of the Amer. Math. Soc. {\bf 16},
        (Providence R.I., 1955).

\bibitem{Grot56} A. Grothendieck, {\em La th\'eorie de Fredholm},
        Bull. Soc. math France, {\bf 84}, 319, (1956).

\bibitem{Isola95} S. Isola, {\em Dynamical zeta functions and
      correlation functions for intermittent interval maps,}
      (Preprint, 1995).

\bibitem{Mayer91} D.H. Mayer, {\em Continued fractions and related
        transformations}, in  Ergodic Theory, Symbolic Dynamics and
        Hyperbolic Spaces. T. Bedford, M. Keane, C. Series (Eds.).
        (Oxford University Press, Oxford, 1991.)

\bibitem{PP} W. Parry and M. Pollicott,
        {\em An analogue of the prime number theorem for closed
         orbits of Axiom A flows},
        {Ann. Math.\bf 118\rm, 573 (1983)}.

\bibitem{PM} Y. Pomeau and P. Manneville, 
        Comm. Math. Phys. {\bf 74}, 189 (1980).
         
\bibitem{PS92} T. Prellberg and J. Slawny,
       {\em Maps of Intervals with Indifferent Fixed-Points -
        Thermodynamic Formalism and Phase-Transitions},
         J. Stat. Phys. {\bf 66}, 503 %-514
        (1992).
\bibitem{Ruelle76} D. Ruelle,
        {\em Zeta functions for expanding maps and Anosov flows,}
         Invent. Math. {\bf 34}, 231 %-242
        (1976).
\bibitem{Ruelle90} D. Ruelle,
         {\em An extension of the theory of Fredholm determinants},
         Math. IHES {\bf 72}, 175-193 (1990).
\bibitem{Shishikura} M. Shishikura,
        {\em The Hausdorff Dimension of the Boundary of the
          Mandelbrot Set and Julia Sets,}
          (Preprint 1991/7) SUNY, Stony Brook (1991).

}
\end{thebibliography}
\end{document}